\documentclass[prc,preprintnumbers,twocolumn,superscriptaddress,amsmath,amssymb]{revtex4}

\usepackage{graphicx}
\usepackage{dcolumn}
\usepackage{bm}
\usepackage{verbatim}
\usepackage{CJK}
\usepackage{braket}
\usepackage{overpic}
\usepackage{rotating}

\begin{document}


\title{Experimentally Study the Deep Dirac Levels with High-Intensity Lasers}

\author{X. P. Zhang}
\affiliation{INPAC, Dept. of Phys.\&Astro., Shanghai Jiao Tong University, Shanghai, 200240, China}

\author{C. B. Fu}
\email[Corresponding author:]{cbfu@sjtu.edu.cn}
\affiliation{INPAC, Dept. of Phys.\&Astro., Shanghai Jiao Tong University, Shanghai, 200240, China}
\affiliation{Shanghai Key Lab for Particle Physics and Cosmology, Shanghai, 200240, China}

\author{D. C. Dai}
\affiliation{INPAC, Dept. of Phys.\&Astro., Shanghai Jiao Tong University, Shanghai, 200240, China}

\date{\today}
\begin{abstract}
Various theories have predicted the deep Dirac levels (DDLs) in atoms for many years. 
However, the existence of the DDL is still under debating, and need to be confirmed experimentally. 
With the development of high intensive lasers, nowadays,  electrons can been accelerated to relativistic energy by high intensive lasers, electron-positron pairs can be created, and nuclear reactions can been ignited, 
which provide a new tool to  explore the DDL related fields.  
In this paper, we propose a new experimental method to study the DDL levels by monitoring nuclei's orbital electron capture life time in plasma induced by high intensive lasers.  
If a DDL exists, a nuclear electron capture rate could be enhanced by factor of over $10^7$, 
which makes it practically detectable in nowadays high intensive laser environments. 
\end{abstract}
\maketitle
\section{Introduction}

Traditionally, in quantum mechanics, when one solves the Schrodinger equations,  Dirac equations, or Klein-Gordon equations, to obtain the bound states of an atom,  at some points, 
the sign of a parameter's square root has to be chosen, and the choice leads either to ``usual" solution or  ``anomalous" solution\cite{QM-Schiff}. 
The traditionally discarded ``anomalous" solutions can be explained as quantum states occupied by  electrons under the ``usual" Bohr ground level. 
The bounding energy is comparable to the rest mass of the electron,
and therefore is called to be deep Dirac level (DDL), electron deep level, or relativistically bounded level in literatures\cite{dombey2006hydrino-theory,naudts2005hydrino-KG-eq}.
The existence of the DDLs are debated theoretically, as well as experimentally.
Some experimental phenomenons  have been explained as existence of the DDLs\cite{phillips2004water2004,mills2003extreme, va2013new}, but being questioned by others\cite{jovicevic2009spectroscopic, rathke2005critical, phelps2005comment, dombey2006hydrino, de2007orthogonality}. 

With the development of high intensity lasers, nowadays, the electrons, and even the nuclei, can be accelerated by the lasers to relativistic energies in a very short distance like 1 $\mu$m, and therefore may populate atoms to the DDL states. 
Due to the very short distance from the electron's DDL orbit to the nuclei, the electron capture (EC) life time can be changed greatly. Therefore one can use the EC rate as an indicator  of the DDL state.
In this paper, we discuss the possibilities of using this novel setup to test the existence of the hypothetical DDLs.

\section{The Deep Dirac Level}

For simplicity, here we only briefly give the DDL solution deduced from the Klein-Gordon equation. The DDL solutions of other equations like the relativistic Schrodinger equation and  Dirac equation can be found in Ref. \cite{paillet2016arguments}.
Considering the Klein-Gordon equation of a hydrogen-like atom\cite{naudts2005hydrino-KG-eq},
\begin{equation} \label{KG-eq.0}
\left[
\left( 
i\hbar\partial_t +U
\right)^2
+\hbar^2 c^2 \Delta 
\right] \Psi_t
= m_0^2 c^4 \Psi_t,
\end{equation}
where $\hbar$ is the Plank constant, $c$ is the speed of light, $Z$ is the charge of the nuclear, $\alpha$ is the fine structure constant, $U(r)=-Z \alpha\hbar c/r$ is the coulomb potential, and $m_0$ is the mass of electron.
This equation has solutions\cite{naudts2005hydrino-KG-eq}: 
\begin{equation}\label{eq.psi.t}
\Psi_t=\frac{R^{s-3/2}  r^{-s} e^{-r/R}}{2^{s-1/2}\sqrt{\pi \Gamma(3-2s)}} e^{-i E t/\hbar},
\end{equation}
where $\Gamma(x)$ is the gamma function, and $s=\frac{1}{2} (1\pm \sqrt{1-4Z^2\alpha^2})$. 
The  corresponding energy $E$ and the orbit $R$ are:
\begin{equation}
E=m_0 c^2\frac{Z\alpha}{\sqrt{s}},
\end{equation}
and
\begin{equation}
R=\frac{\hbar}{m_0c}\frac{1}{\sqrt{s}}.
\end{equation}

The $s=\frac{1}{2} (1- \sqrt{1-4Z^2\alpha^2}) \simeq 0$ solution is normal non-relativistic one, i.e., the Bohr state. 
In this case, the corresponding energy, orbit, and wave function are,
\begin{equation}
E_0=m_0 c^2(1-\frac{1}{2}Z^2\alpha^2 + ...)\simeq m_0 c^2-Z^2\cdot13.6 \,\, eV, 
\end{equation}
\begin{equation}
r_0\simeq\frac{\hbar}{m_0c}\frac{1}{Z \alpha}\simeq \frac{0.53}{Z}\, \overset{\circ}{A},
\end{equation}
and 
\begin{equation}\label{eq.psi.0}
\psi_t\simeq\frac{ e^{-r/r_0}}{r_0^{3/2}\sqrt{\pi}} e^{-i E_0 t/\hbar}.
\end{equation}

The $s=\frac{1}{2} (1+ \sqrt{1-4Z^2\alpha^2})\simeq 1$ solution is the ``anomalous'' one, i.e. the DDL level.
In this case, the energy, orbit, and wave function can be simplified as,
\begin{equation}
E_0^\#=m_0 c^2\cdot Z\alpha\simeq m_0 c^2-(511-3.72\cdot Z) keV,
\end{equation}
\begin{equation}
r_0^\#\simeq\frac{\hbar}{m_0c}\simeq 0.0039\, \overset{\circ}{A}.
\end{equation}
\begin{equation}\label{eq.psi.ddl}
\psi_t^\#\simeq\frac{ e^{-r/r_0^{\#}}}{r\sqrt{2\pi r_0^{\#}}} e^{-i E_0^\# t/\hbar},
\end{equation}
As one can see, taking $Z=1$ as an example, 
in the case of the DDL, the electron is deeply bounded to $0.5073$~MeV, 
compared with the well-known Bohr case $13.6$~eV.
 The DDL's orbit is only  about 390 fm away from the nucleus, 
 compared with the Bohr orbit, $0.53\ \overset{\circ}{A}$. 

Because the wave function $\lim\limits_{r \to 0}\psi_t^{\#} =\infty$,  this ``un-physical" solution was traditionally rejected in quantum mechanic textbooks\cite{QM-Schiff}.
The infinity comes from the assumption that the nucleus is point-like, 
and therefore the Coulomb potential is infinite at r = 0.
Even through, the wave function $\psi_t^{\#}$  is still square integrable, i.e. $\int_0^\infty |\psi_t^{\#}|^2 4\pi r^2 {\rm d}r $ is not infinity. In fact, the wave function Eq. \ref{eq.psi.t} has already been normalized to $1$ in $r\in[0, \infty)$.

\section{populate electrons to the DDL}\label{sec.pop}

Directly populating DDLs  via photoemission must be highly forbidden, 
because otherwise a lot of high energy ambient photons can be observed 
due to the fact that the DDL is about 0.5 MeV below the normal Bohr ground state. 

The DDLs may be populated via electron-positron pair effect. 
When a relativistic electron approach a nucleus, $e^-e^+$ pairs can produced
through the following two processes:
\begin{equation}
Z+e^- \rightarrow Z+ 2 e^- +e^+
\end{equation}
\begin{equation}
Z+e^- \rightarrow Z+e^-+\gamma\rightarrow Z+ 2 e^- +e^+
\end{equation}
The larger the electrical field, i.e. the closer to the nuclear, the higher the possibility is.
Therefore, the electrons in the $e^+e^-$ pairs produced near the nuclei have higher chance to be bounded to be DDLs.
Because the electron in the DDL is very closer to the nucleus, it has higher possibility to be caught which results in a short EC life time, if the EC decay model is allowed. 
The changing of the nuclear EC life time may be used as an indicator of the DDLs. 

With the development of high intensive laser technologies, 
the intensities of today's laser could be as high as $10^{22}$ W/cm$^2$\cite{RevModPhys.84.1177}. 
With high intensive lasers, the electron-positron pairs have been experimentally observed\cite{e-e+PhysRevLett.106.035001}. 
When an $e^+e^-$ pair is created near an nuclei, the positron escapes due to the coulomb field, and the electrons may be caught to the DDLs.

The DDL may also be produced through the mechanism called Nuclear Excitation by Electron Transition (NEET)\cite{NEET-PhysRevC.88.054616}.
When an electron moves from an outside orbit to an inner one, the energy difference $\Delta E_e$ between the two orbits  is usually carried away by the $X$-ray or the Auger electron. 
However, it is possible that part of that energy can be transferred to the nucleus and get it  excited. 
That is why it got its name, the nuclear excitation by electron transition. 
The NEET have been founded in several nuclei, e.g., $^{197}$Au, $^{189}$Os, and $^{193}$Ir etc.
\cite{Au197-NEET-PRL85.1831,Os189-PRC61.051304} 
With the high intensive lasers, the DDL may be produced through the NEET mechanism.  

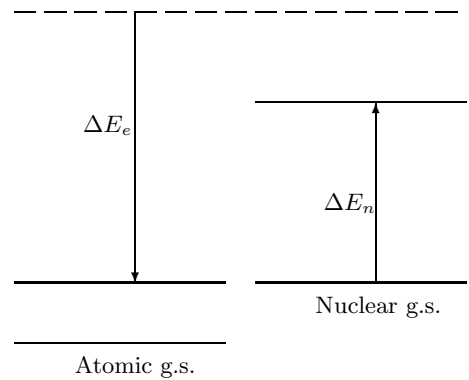
\begin{figure}
\begin{center}
\setlength{\unitlength}{8cm}
\begin{picture}(1,0.61)
	\put(0.1,0.05){\line(1,0){0.35}}
	\put(0.1,0.15){\line(1,0){0.35}}
	\put(0.3,0.6){\vector(0,-1){0.45}}
	\put(0.2,0){Atomic g.s.}
	\put(0.2,0.4){ $\Delta E_e$}
	\multiput(0.1,0.6)(0.05,0){15}{\line(1,0){0.035}}

	\put(0.5,0.15){\line(1,0){0.35}}
	\put(0.5,0.45){\line(1,0){0.35}}
	\put(0.7,0.15){\vector(0,1){0.3}}
	\put(0.6,0.10){Nuclear g.s.}
	\put(0.6,0.27){ $\Delta E_n$}
\end{picture}
\caption{The energy scheme of the NEET. 
When an electron falls into an orbit of an atom, part of the energy difference  between the two orbits $\Delta E_e$ may be transferred to the nucleus, and then boosts the nucleus to its excited state $\Delta E_n$. 
}
\label{fig.wavefunciton}
\end{center}
\end{figure}

In a typical high-intensity laser experiment, a laser pulse's duration is normally shorter than nanosecond. 
The radiation produced in this ns time interval include $X$-rays, $\gamma$-rays, neutrons, electrons, positrons, ions, and  other radioactive isotopes. 
The photons (X-ray and $\gamma$-ray) decline immediately in nanoseconds.
After the original pulse, the high energy electrons, as well as the ions,  fly away from the target, 
and hit on materials around  to have Bremsstrahlung $\gamma$-ray or other secondary neutrons.
This process may last another several nanoseconds depending on the experimental setups.
The positrons may last longer than tens ns, but then their annihilation energy must be smaller than $<2m_0c^2\simeq1.022$ MeV.

From the experimental point of view, the positrons themselves only  are hardly to be used as the existing signs of the DDLs. 
After all, a lot of positrons, as well as different energy photos, are produced by lasers in a time interval smaller than one nanosecond. It's very hard to trace the history of a positron experimental in this circumstance.

There are two kinds of $\gamma$ background in the typical high-intensity laser experiment: laser induced and ambient gamma.
The later one comes from cosmic rays or the radiation of the ambient materials.
The laser background  almost completely disappears after roughly about 100 ns.
The ambient $\gamma$ background normally is very small compared with the laser $\gamma$ background at the beginning,
but it will dominate after about $t\gtrsim 1$ min. 
Therefore,  a window roughly from 100 ns to 1 min is ideal for the DDL detecting.
 
After about 100 ns, almost all observed photons with $E_\gamma>1.1$ MeV must come from 5 sources: 
neutron capture reactions,
activated radioactive isotopes, 
ambient sources,
$e^+$ annihilation,
and the possible EC from DDLs.
The first two sources have their own characteristic $\gamma$-ray, and can be distinguished from the DDL cases.
The ambient background can be compressed by choosing a proper time window, let's say, $t\in[10^2,10^{11}]$ns.
The positron background can be compressed by choosing a proper energy window, $E_\gamma>$1.1 MeV.
Base on the discussion above, the laser-induce EC decay may be an ideal probe for the DDL studies.


\section{Electron Capture Rate of the DDL}

Let's consider the EC process, 
\begin{equation}
^A_Z X+e^- \rightarrow\, ^{\,\ \ A}_{Z-1}Y+\nu_e.
\end{equation}
The EC decay probability per time unit is given by Fermi's golden rule,
\begin{equation}
\lambda=\frac{2\pi\,\rho_f}{\hbar}|\Bra f \hat{O}  \Ket i |^2,
\end{equation}
where $i$, $f$ are initial and final states respectively, $\rho_f$ is the neutrino final states per energy unit, and $\hat{O}$ is the weak interaction operator.

 There are one proton, one electron, and the other nucleons in the initial state, $^A_Z X$, 
i.e. $\ket{ i } = \ket{p,e }\ket{ ^{A-1}_{Z-1}Y}$.
There are one neutron, one neutrino, and the other nucleons in the finial state, $^{\,\,\,\,\,\,A}_{Z-1}Y$, 
i.e. $\ket{f}=\ket{n,\nu}\ket{^{A-1}_{Z-1}Y}$.
Since the operator $\hat{O}$ acts only on weak-interaction participants,  
the EC decay rate $\lambda$ is roughly proportional to  the probability of finding the electron in the nuclear volume\cite{RMP-Review-EC-bambynek1977orbital}
, i.e.
\begin{equation} \label{eq.int.psi.rn}
\lambda\propto \int_0^{r_n} |\psi_e |^2 \cdot 4\pi r^2 {\rm d}r,
\end{equation}
where $\psi_e$ is the electron's wave function, and $r_n$ is radius of the nucleus.
Here we use the assumption that the nucleon's wave function is constant in the nuclear volume. 
Furthermore, the coulomb interaction difference  between the normal Bohr state and the DDL is very small compared with the strong interaction inside the nuclei. 
Therefore, the other parts of the matrix for the Bohr level and the DDL are roughly same.
Their EC decay ratio can be simplified as
\cite{EC-Cr-Section-bahcall1963exchange, RMP-Review-EC-bambynek1977orbital}:
\begin{eqnarray}
\frac{T_{1/2}^{\#}}{T_{1/2}} &\simeq&
\left[ 
\frac{Q}{Q^{\#}}
\frac{|\psi_t(r_n)|}{|\psi_t^{\#}(r_n)|}
\right]^2 \\
&\simeq&
\left[ 
\frac{1}{1-m_0 c^2/Q}
\cdot
\frac{|\psi_t(r_n)|}{|\psi_t^{\#}(r_n)|}
\right]^2 \label{eq.rr.rate}
\end{eqnarray}
where the $Q$ and $Q^{\#}$ is the reaction Q-value for the Bohr state and the DDL respectively,
the $\frac{|\psi_t(r_n)|}{|\psi_t^{\#}(r_n)|}$ is the ratio of the electron wave functions evaluated at the nuclear surface $r=r_n$.
Here we take
\begin{equation}
Q^{\#}\simeq Q -m_0 c^2.
\end{equation}

From Eq.\ref{eq.psi.0} and \ref{eq.psi.ddl}, we have
\begin{equation}
\frac{|\psi_t(r_n)|}{|\psi_t^{\#}(r_n)|}
=\frac{\sqrt{2}(r_0^{\#})^{1/2}r_n}{r_0^{3/2}}\exp[{r_n/r_0^{\#}-r_n/r_0}].
\end{equation}
Since $r_n$ is in order of fm level, $r_0^\#$ is about 390 fm, and $r_0$ is about 0.53 $\overset{\circ}{A}$, we have $\exp[{r_n/r_0^{\#}-r_n/r_0}]\simeq 1$.
The equation can be re-written as,
\begin{equation}
\frac{|\psi_t(r_n)|}{|\psi_t^{\#}(r_n)|}
\simeq \frac{r_n}{r_0} \sqrt{2Z\alpha},
\end{equation}
Insert it to the Eq.\ref{eq.rr.rate}, we have
\begin{eqnarray}
\frac{T_{1/2}^{\#}}{T_{1/2}} &\simeq&
\frac{2Z\alpha}{(1-m_0 c^2/Q)^2}
\left(\frac{r_n}{r_0}\right)^2.
\end{eqnarray}
Some EC decay nuclei's life time are shown as $T_{1/2}^{(DDL1)}$ in Tab. \ref{tab.EC.exp}.

\begin{figure}
\begin{center}
\begin{overpic}[width=8cm]{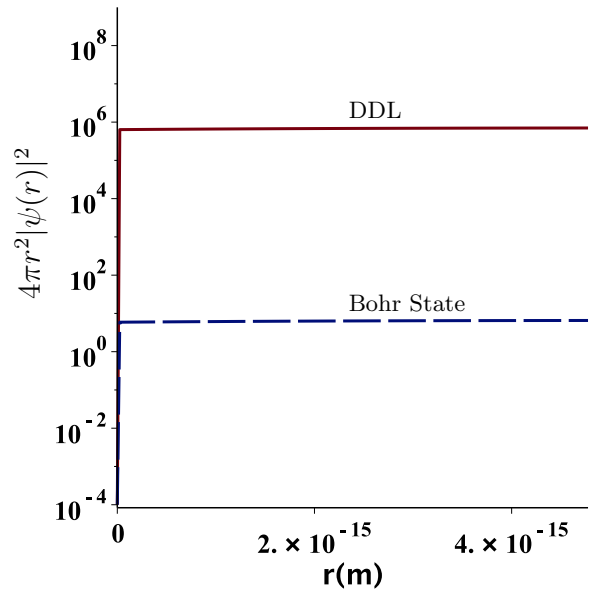}
	\begin{sideways}
	    \put(50,0){\large $4\pi r^2 |\psi(r)|^2$}
	\end{sideways}
	\put(50,80){DDL}
	\put(50,48){Bohr State}
\end{overpic}
\caption{A comparison of the numerical evaluation of $4\pi r^2 |\psi(r)|^2$ ($r<r_n$) for the $^{62}$Cu's Bohr state (dash line) and DDL(solid line). }
\label{fig.wavefunciton}
\end{center}
\end{figure} 

We also numerically solve the Eq. \ref{KG-eq.0} with a more realistic potential, 
specifically, assuming  that the charge is evenly distributed in the nucleus. 
The  potential $U$ in Eq. \ref{KG-eq.0} has the form:
\begin{equation}
 U(r)=\left\{
\begin{array}{l l l}
-\frac{Z \alpha\hbar c}{r}       &      & {r      >      r_n}\\
 -\frac{Z \alpha\hbar c\,r^2}{r_n^3}       &      & {r      \le      r_n}
\end{array} \right. 
\end{equation}
Once obtaining the numerical wave functions, the EC decay ratio is then calculated according to Eq. \ref{eq.int.psi.rn}. 
The results are shown as $T_{1/2}^{(DDL2)}$ 
in Tab. \ref{tab.EC.exp}.
As an example, the numerically-solved wave functions for $^{62}$Cu's Bohr state and  DDL are shown in Fig. \ref{fig.wavefunciton}.

The nuclei listed in the Tab.\ref{tab.EC.exp} can be created from stable nuclear  one-nucleon-transferring reactions. 
Therefore, they are relatively easy to be obtained by using high intensity laser beams. 
Furthermore, because their life time, $T_{1/2}^{(DDL)}$, are much longer than 100 ns, 
but relatively shorter regarding the cosmic and ambient radiation backgrounds (look the argument in Sec.\ref{sec.pop}), 
a high signal-to-noise ratio can be achieved if these nuclei are used. 
In Ref. \cite{ZXP2017}, the gamma ray intensity with a decay life-time of about 300 $\mu$s
may relate to the DDLs of $^{53}$Fe and/or $^{62}$Cu.

\begin{table}[t]
\begin{ruledtabular}
\caption {The new life time of several radioactive nuclei if the DDL exist. 
Here the nuclear radius of $r_n=1.2\cdot A^{1/3}$[fm] is taken in the calculations. } 
\label{tab.EC.exp}
\begin{tabular}{c c c r c  c}   
Nucleus & $Q$ (MeV) & Z & $T_{1/2}^{(0)}$& $T_{1/2}^{(DDL1)}$ & $T_{1/2}^{(DDL2)}$ \\
\hline
$^{7}$Be & 0.861 & 4 & 53.2 d         & 21 ms        &52 ms\\
$^{11}$C & 1.982 & 6 & 20.3 m    & 18 $\mu$s &19 $\mu$s\\
$^{13}$N & 2.220 & 7 & 10.0 m     & 14 $\mu$s &16 $\mu$s \\
$^{15}$O & 2.757 & 8 & 122 s          & 4.3 $\mu$s   & 4.8 $\mu$s \\
$^{23}$Mg & 4.056 & 12 & 11.3 s     & 1.5 $\mu$s   &1.9 $\mu$s \\
$^{30}$P & 4.232 & 15 & 2.50 m    & 47 $\mu$s &64 $\mu$s\\
$^{53}$Fe & 3.742 & 26 & 8.51 m  & 1.3 ms          &2.9 ms   \\
$^{62}$Cu & 3.958 & 29 & 9.67 m  & 2.2 ms          &5.9 ms    \\
$^{63}$Zn & 3.367 & 30 & 38.47 m & 10 ms       &30 ms   \\
$^{64}$Cu & 1.675 & 29 & 12.7 h      & 0.28  s          &0.7  s\\
\end{tabular} 
\end{ruledtabular}
\end{table}


\section{Summary}

In this work we provide a new experimental method to explore the deep Dirac levels.
The DDLs may be populated by high intensive lasers through the mechanism of $e^+e^-$ pair or the NEET, 
Because the DDL orbit is very close to the nucleus, the electron capture  (EC) rate can be enhanced greatly.
We estimate that the EC rate will be about $10^8$ times higher if the DDL exist.
The characteristic EC decay $\gamma$-ray could be used as the indicator of the DDL's existence. 
We suggest a detecting time window, from 100 ns to 1 min, 
which can avoid both the laser-induced $\gamma$-ray at $t=0$ and the ambient $\gamma$-ray.
We also provide  several  nuclei candidates which can be relatively easily created in laser setups. 
Their relatively high decay energies can benefit the detecting signal-to-noise ratios.  
We expect  that this new laser-induced EC decay method will help to understand more about the long-existing DDL puzzle.

This work is supported by 
the National Nature Science Foundation of China (Grant Nos. 11375114), 
One of us (CBF) thanks Shanghai Municipal Science and Technology
Commission for the supports (under grant No. 11DZ2260700)).


\end{document}